\documentstyle[epsf]{mn}

\title[]{{\sl XMM-Newton} observations of the polars EV UMa, RX
J1002-19 and RX J1007-20\thanks{Based on observations
obtained with XMM-Newton, an ESA science mission with instruments
and contributions directly funded by ESA Member States and the USA
(NASA).}}

\author[Ramsay \& Cropper]{
Gavin Ramsay and Mark Cropper\\
Mullard Space Science Lab, University College London,
Holmbury St. Mary, Dorking, Surrey, RH5 6NT, UK\\}
\date{Received: }

\begin{document}
\outer\def\gtae {$\buildrel {\lower3pt\hbox{$>$}} \over 
{\lower2pt\hbox{$\sim$}} $}
\outer\def\ltae {$\buildrel {\lower3pt\hbox{$<$}} \over 
{\lower2pt\hbox{$\sim$}} $}
\newcommand{\ergscm} {ergs s$^{-1}$ cm$^{-2}$}
\newcommand{\ergss} {ergs s$^{-1}$}
\newcommand{\ergsd} {ergs s$^{-1}$ $d^{2}_{100}$}
\newcommand{\pcmsq} {cm$^{-2}$}
\newcommand{\ros} {\sl ROSAT}
\newcommand{\exo} {\sl EXOSAT}
\newcommand{\xmm} {\sl XMM-Newton}
\def\rchi{{${\chi}_{\nu}^{2}$}}
\newcommand{\Msun} {$M_{\odot}$}
\newcommand{\Mwd} {$M_{wd}$}
\def\Mdot{\hbox{$\dot M$}}
\def\mdot{\hbox{$\dot m$}}

\maketitle

\begin{abstract}

We present {\xmm} data of three strongly magnetic cataclysmic
variables (polars) EV UMa, RX J1002-19 and RX J1007-20. These include
the polar with the shortest orbital period (EV UMa) and the polar with
one of the highest magnetic field strengths (RX~J1007--20). They
exhibit a range of X-ray spectral characteristics which are consistent
with their known magnetic field strength. We find that two of the
systems show evidence for an absorption dip in soft
X-rays. Their profiles are well defined, implying that the stream is
highly collimated. We determine the mass transfer rate for the two
systems with known distances.  We determine that the mass of the white
dwarf in EV UMa and RX J1007-20 is $\sim$1\Msun while in RX J1002-19
it is closer to $\sim$0.5\Msun.

\end{abstract}


\section{Introduction}

Polars or AM Her systems are accreting binary systems in which
material transfers from a dwarf secondary star onto a magnetic
($B\sim$10--200MG) white dwarf through Roche lobe overflow. At some
height above the photosphere of the white dwarf a shock forms. Hard
X-rays are generated in this post-shock flow ({\it c.f.} Wu 2000 for a
review of the shock structure in these systems). Cyclotron radiation
is also emitted in this post-shock flow by electrons spiralling around
the magnetic field lines: this radiation is emitted in the optical
band. Some fraction of the hard X-rays and cyclotron emission are
intercepted by the photosphere of the white dwarf and are re-emitted
at lower energies. Soft X-rays can also be produced by dense `blobs'
of material which impact directly into the white dwarf.

As part of a programme to determine how the balance of soft and hard
X-rays are affected by system parameters such as the magnetic field,
we have observed a number of polars using {\xmm}. These include DP
Leo, WW Hor (Ramsay et al 2001), BY Cam (Ramsay \& Cropper 2002a) and
CE Gru (Ramsay \& Cropper 2002b). Here, we present the results on
three further polars, EV UMa, RX~J1002--19 and RX~J1007--20, all of
which were discovered using {\ros}. We show their main system
parameters in Table \ref{parameters}. These systems show a range of
parameters, including the polar with the shortest orbital period (EV
UMa), and a polar with one of the highest magnetic field strengths 
(RX~J1007-20). In this paper we first discuss their temporal properties
and then their spectral properties.

EV UMa (RE~J1307+535) also has exhibited the highest recorded degree
of polarisation in any polar (or any astrophysical object for that
matter) when the circular polarisation varied from +50 to --20 percent
over an orbital cycle when it was in an intermediate accretion state
(Hakala et al 1994). The degree of polarisation was reduced when it
was at higher accretion states (presumably because of the increased
dilution of the polarisation by a bright accretion stream) and the
system has a significantly different light curve (Katajainen et al
2000). Although this object has been relatively well observed in the
optical and also the EUV (using the Wide Field Camera on {\ros},
Osborne et al 1994) no X-ray observations of this object have been
published.

RX J1007--20 was shown to show a prominent dip in the soft X-ray light
curve, which was attributed to the accretion stream obscuring the
emission from one of the accretion sites and also has a very high
soft/hard X-ray ratio (Reinsch et al 1999). 

Very little information has been published on RX J1002~--19.

\begin{table}
\begin{center}
\begin{tabular}{rrr}
\hline
Source & Orbital Period & Magnetic field \\
\hline
EV UMa & 79.69m$^{1}$ & 30--40 MG$^{1}$\\
RX J1002-19 & 107m$^{2}$ & \\
RX J1007-20 & 208m$^{2}$ & 92MG$^{3}$\\
\hline
\end{tabular}
\end{center}
\caption{The main system parameters of EV UMa, RX J1002-19 and RX
J1007-20. (1) Osborne et al 1994, (2) Beuermann \& Burwitz 1995, (3)
Reinsch et al 1999.}
\label{parameters}
\end{table}

\section{Observations}

The satellite {\xmm} was launched in Dec 1999 by the European Space
Agency. It has the largest effective area of any X-ray satellite and
also has a 30 cm optical/UV telescope (the Optical Monitor, OM: Mason
et al 2001) allowing simultaneous X-ray and optical/UV coverage. The
EPIC instruments contain imaging detectors covering the energy range
0.1--10keV with moderate spectra resolution. The start time of the
EPIC MOS observations preceded the EPIC pn observations. The OM data
were taken in one optical band ($V$ band) and two UV filters (UVW1:
2400--3400 \AA, UVW2: 1800--2400 \AA) (in that sequence). The
observation log is shown in Table \ref{log}. The data obtained using
the RGS (den Herder et al 2001) were of low signal-to-noise and are
therefore not discussed further.

The data were processed using the {\sl XMM-Newton} {\sl Science
Analysis Software} v5.3. For the EPIC pn detector (Str\"{u}der et al
2001), data were extracted using an aperture of 40$^{''}$ centered on
the source. Background data were extracted from a source free
region. For the EPIC MOS detectors (Turner et al 2001) we extracted
the background from an annulus around the source. The background data
were scaled and subtracted from the source data. The OM data were
analysed in a similar way using {\tt omichain} and {\tt omfchain}
(this latter task was not incorporated into SAS v5.3 but has recently
been added to SAS v5.3.3). Data were corrected for background
subtraction and coincidence losses (Mason et al 2001). In extracting
the EPIC pn spectrum, we used single and double pixel events and used
the response files epn\_ff(lw)20\_sdY9\_thin.rmf for the full frame
and (large window) modes respectively. In the case of the MOS data we
used the response files m[1-2]\_thin1v9q19t5r5\_all\_15.rsp.

We show in Table \ref{log} the mean $V$ mag determined using the OM
for each source at the time of our observation. EV UMa was seen by
Osborne et al (1994) in two different brightness levels: $V=$17--18
and $V=$20--21. Katajainen et al (2000) found it to be in an even
higher accretion state at $V$=16--17. EV UMa was therefore in a high
accretion state during our {\xmm} observations.

RX J1002-19 is quoted as $V\sim$17 by Beuermann \& Burwitz (1995) and
a {\ros} PSPC count rate of 0.63 ct/s. Our detection of $V\sim$19.6
was made during the X-ray faint phase, so that is probably an
underestimate of its mean orbital brightness. However, this together
with its mean X-ray brightness ({\it c.f.} Figure \ref{lightrx1002})
suggests that RX J1002-19 was in an intermediate accretion state.  

RX J1007-20 was identified with a $V\sim$18 star by Thomas et al
(1998) and had a peak {\ros} PSPC count rate of $\sim$1 ct/s. This
suggests that it was in a similarly high accretion state at the epoch
of the {\xmm} observations.

\begin{table*}
\begin{center}
\begin{tabular}{lrrr}
\hline
 & EV UMa & RX J1002-19 & RX J1007-20\\
\hline
Date & 2001 Dec 8 & 2001 Dec 10 & 2001 Dec 7\\
EPIC MOS & LW thin 7772 sec & FF thin 5992 sec & LW thin 7507 sec\\
EPIC pn & LW thin 5468 sec & FF thin 3668 sec& LW thin 5203 sec\\
RGS & 8230 sec & 6455 sec & 7965 sec\\
OM & Image/fast UVW1 1500 sec & Image/fast UVW1 1500 sec &Image/fast
 UVW1 1500 sec\\
OM & Image/fast UVW2 3900 sec& Image/fast UVW2 2000 sec&Image/fast
 UVW2 3600 sec\\
OM & Image/fast V 1500 sec& Image/fast V 1500 sec& Image/fast V 1500 sec\\
Mean $V$ mag & 16.6 & 19.6 & 18.5 \\
Accretion state & High & Intermediate & High\\
\hline
\end{tabular}
\end{center}
\caption{The log of {\xmm} observations of EV UMa, RX J1002-19 and 
RX J1007-20. LW refers to large window mode and FF full frame
mode, while thin refers to the filter used. The exposure time in each
detector is shown in seconds. The 
UVW1 filter has a coverage 2400--3400 \AA\hspace{1mm} and UVW2 1800--2400 \AA.}
\label{log}
\end{table*}

\section{Light curves}

\subsection{EV UMa}

\subsubsection{XMM-Newton data}

\begin{figure}
\begin{center}
\setlength{\unitlength}{1cm}
\begin{picture}(8,10)
\put(-0.5,-1.){\includegraphics{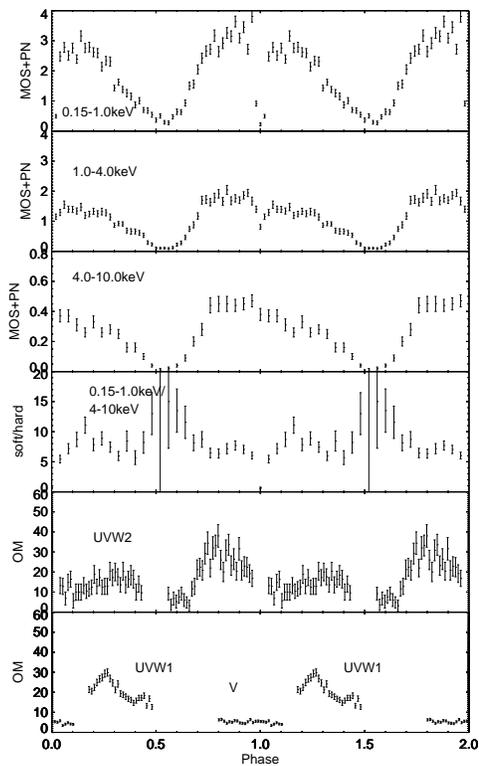}}
\end{picture}
\end{center}
\caption{The phased and binned X-ray and optical light curves of EV
UMa. We have folded the data on the period of Osborne et al (1994) and
defined phase zero as the center of the absorption dip. Energy bands
0.15--1.0keV and 1.0--4.0keV have bin widths of 0.02 cycles, while the
energy band 4.0-10.0keV has a bin width of 0.04 cycles. The optical
data were binned into 60sec bins (0.012 cycles). The units for the OM
data are $10^{-16}$ \ergscm.}
\label{lightev} 
\end{figure}

Both the EPIC pn and MOS data covered a full orbital cycle. We
therefore folded the data from each detector on the orbital period and
then binned and co-added them. The X-ray light curve shown in Figure
\ref{lightev} has a prominent bright phase lasting over half the
orbital cycle with a broad minimum where the flux is much reduced. In
the center of the bright phase there is a prominent dip which is
strongest at soft energies: this is characteristic of photo-electric
absorption and is due to the accretion stream obscuring the X-rays
emitted from one of the accretion regions on the white dwarf. In
harder X-rays, the flux after the dip is significantly lower compared
to the flux before the dip. We go on to discuss this further below.

Our UVW2 data covers a large fraction of the orbital period
($\sim$0.8). We find that it increases in flux at the same phase at
the rise to the X-ray bright phase. Unfortunately the data do not
cover the phase of the accretion stream dip. After the dip, the flux
is around half that of the bright phase before the dip. The $V$ band
data shows no significant variability, while the short section of UVW1
data shows an initial rise followed by a decline.

\subsubsection{ROSAT data}

{\ros} made two pointed observations of EV UMa, but on both occasions
it was in a low accretion state and not detected.  Osborne et al
(1994) presents optical photometry and EUV data taken from the {\ros}
all sky survey using the wide field camera (WFC). The phase-folded
optical data shows a basically sinusoidal shape with a secondary
minimum (which is most prominent in red light) occurring at what would
otherwise be maximum brightness.  The WFC S1 filter data
(0.09-0.21keV) shows a bright phase lasting $\sim$0.4--0.5 cycles with
the remainder being consistent with zero flux: maximum flux coincides
with the bright phase in the optical band.

\begin{figure}
\begin{center}
\setlength{\unitlength}{1cm}
\begin{picture}(8,4.)
\put(-1.5,-15.){\includegraphics{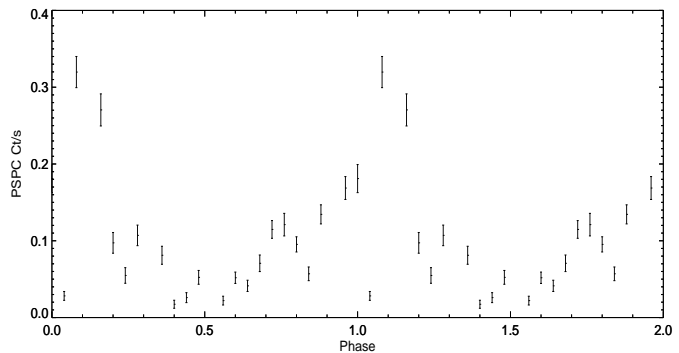}}
\end{picture}
\end{center}
\caption{The {\ros} all sky survey data of EV UMa folded on the
ephemeris of Osborne et al (1994).}
\label{rass} 
\end{figure}

Osborne et al (1994) do not show the {\ros} X-ray all sky survey
data. We have extracted those data from the archive at MPE and
analysed the data using {\tt ftool} procedures and folded the data on
the ephemeris of Osborne et al (1994) whose time zero is the maximum
of optical pulse. Based on the estimated errors of this ephemeris we
expect there to be an uncertainty of 0.05 cycles in the phasing of the
{\ros} data as in the case of Osborne et al (1994). We show the folded
and binned light curve in Figure \ref{rass}. The light curve shows a
marked degree of irregularity. Some of this is probably attributable
to that the fact that we have not taken the effects of vignetting into
account. However, the light curve clearly shows a broad general
variation similar to our {\xmm} data and a narrow dip near
$\phi$=0.0. This implies that the secondary minimum in the optical is
phased with the general X-ray maximum and the narrow absorption dip at
$\phi$=0.0. The EUV data shows no evidence for such an absorption dip,
but this may be due to the relatively poor sensitivity of the WFC.

\subsubsection{Phasing the optical and X-ray data}
\label{evphase}

We now go on to relate our {\xmm} observations with those made
previously and in particular the optical and X-ray light curves.  We
find that the dip in the {\xmm} data occurs at $\phi$=0.65 on the
ephemeris of Osborne et al (1994) which is based on optical data taken
in 1992.  The dip seen in the {\ros} X-ray light curve (taken in 1990)
occurs at $\phi\sim$0.05 (Figure \ref{rass}).  This implies that the
accumulated phase drift between the {\ros} and the {\xmm} data is
$\sim$0.4 cycles.

The Hakala et al (1994) $R$ band data (taken in 1994), can therefore
be phased such that the positive circular polarisation peak coincides
with the absorption dip at $\phi$=0.0 (Figure \ref{lightev}).  Also,
the data of Katajainen et al (2000) has to be advanced in phase by
$\phi\sim$0.2. This places their narrow dip-like feature in $UBV$
bands and the maximum negative circular polarisation (at
$\phi\sim$0.20-0.25 on the ephemeris of Osborne et al 1994) at the
same phase as the X-ray dip in our {\xmm} data.  Consequently, we find
that we expect maximum negative circular polarisation at the phase of
maximum brightness whereas Hakala et al found it to be maximally
positively polarised. It is possible that there was a reversal in the
sign of the polarisation, but it is much more likely that there was an
incorrect calibration in either the Hakala et al (1994) or Katajainen
et al (2000) data. The data of Hakala et al (1994) were taken using a
CCD polarimeter which had an arbitrary beam orientation and there may
have been difficulties in determining the correct sign for the
circular polarisation (Hakala, priv comm).

Our {\xmm} X-ray data indicate that there is almost a complete absence
of X-ray flux at the expected phase at which the secondary accretion
region of Hakala et al (1994) would be most prominent. This calls into
question the existence of this second region. The high inclination of
this system (Hakala et al 1994) would allow the cyclotron emission
from the prime accretion region to be viewed from beneath while the
foot-point of the magnetic field was behind the limb of the white
dwarf. This could be the explanation for the circular polarisation
sign reversal.

Turning to the UV light curves (Figure \ref{lightev}), the low flux
levels in the UVW2 data near $\phi\sim$0.6, and the low X-ray flux,
corresponds to the prime accretion region moving behind the limb of
the white dwarf. The broad depression around $\phi$=0.0 may be due to
the same absorbing material that causes the narrow dip seen in soft
X-rays.

\subsection{RX J1002-19}

\begin{figure}
\begin{center}
\setlength{\unitlength}{1cm}
\begin{picture}(8,9)
\put(-0.5,-1.7){\includegraphics{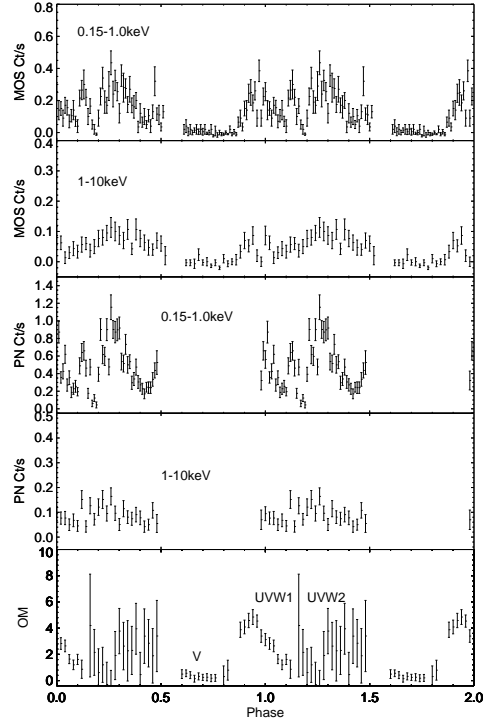}}
\end{picture}
\end{center}
\caption{The binned X-ray and OM light curves of RX J1002-19. We fold
the data on the orbital period of Beuermann \& Burwitz (1995). The
phasing is arbitrary. The 0.15--1.0keV data have been binned into 0.01
cycle bins while the 1.0--10.0keV band data and the OM data have been
binned into 0.02 cycle bins. The units for the OM data are
$10^{-16}$ \ergscm.}
\label{lightrx1002} 
\end{figure}

The X-ray and optical light curve of RX J1002-19 is shown in Figure
\ref{lightrx1002}. This is the first published X-ray light curve of
this source. The orbital coverage is almost complete in the MOS
instruments ($\Delta\phi\sim$0.9) with a bright phase lasting
$\Delta\phi\sim$0.65-0.70 cycles and a faint phase where the flux is
essentially zero. This suggests that the accretion region causing the
X-ray flux is in the upper hemisphere of the white dwarf.

We also find the presence of a structured dip in the soft X-ray light
curve which is not present in the harder band at $\phi\sim$0.18. This
suggests that the dip is due to absorption of X-rays from the
accretion region by the accretion stream as found in EV UMa. The UVW1
data shows the appearance of the X-ray region coming into view at the
same time as the start of the X-ray bright phase. The flux then
decreases around $\phi$=0.97. It is possible that this decrease maybe
related to the absorption dip seen in soft X-rays at $\phi\sim$0.18.
The $V$ band data show a low flux level during the faint X-ray phase.

\subsection{RX J1007-20}

\begin{figure}
\begin{center}
\setlength{\unitlength}{1cm}
\begin{picture}(8,9)
\put(-0.5,-1.7){\includegraphics{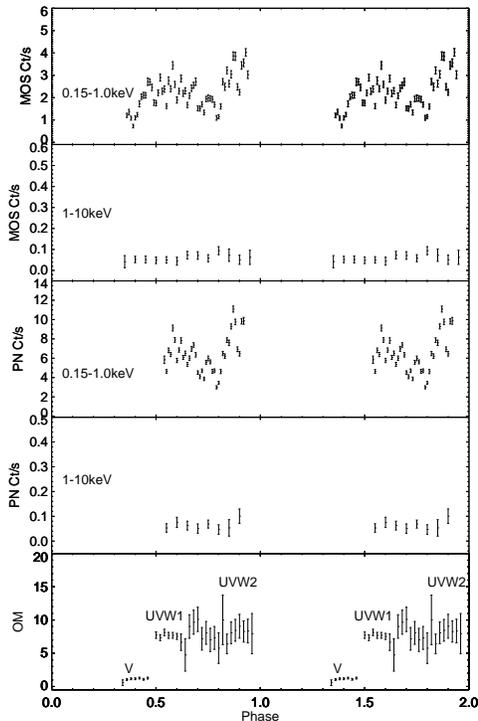}}
\end{picture}
\end{center}
\caption{The binned X-ray and OM light curves of RX J1007-20. We fold
the data on the orbital period of Beuermann \& Reinsch (1995). The
phasing is arbitrary. The 0.15--1.0keV data has been binned into 0.01
cycle bins and the 1.0--10.0keV data into 0.05 cycle bins. The OM data
has been binned into 0.02 cycle bins. The units for the OM data are
$10^{-16}$ \ergscm.}
\label{lightrx1007} 
\end{figure}

Our X-ray data covers approximately half of the binary orbit (Figure
\ref{lightrx1007}). By comparison with both EV UMa and RX J1002-19 the
flux in the 1-10keV energy band is very low. In contrast, below 1keV,
it is bright and highly variable. The {\ros} light curve (Reinsch et
al 1999) also shows a variable light curve with a strong dip seen in
soft X-rays (no counts at the dip maximum) which they interpreted as
an absorption dip similar to that seen in our observations of EV UMa
and RX J1002-19.  However, such a strong dip is not seen in our data
and therefore it is probable that the dip occured at an orbital phase
which we did not observe. (The orbital period is not sufficiently well
known to co-phase the {\xmm} data with the archived {\ros}
data). Reinsch et al (1999) concluded that one pole was visible
throughout the orbital cycle although is it close to the limb of the
white dwarf at certain phases.

\section{X-ray spectra}

\subsection{The spectra}

In two of the three polars, we extracted spectra from a subset of the
available data. In the case of EV UMa (Figure \ref{lightev}) there was
some evidence that two distinct emission regions are present. We
therefore extracted spectra from $\phi$=0.70--0.95 and
$\phi$=0.05--0.35. These phase ranges excluded the deep absorption
dip. For RX J1002--19 we excluded data from the faint phase and also
the phase interval which included an absorption dip (centered on
$\phi$=0.18, Figure \ref{lightrx1002}). We did not exclude any of the
RX J1007--20 data.

\begin{figure}
\begin{center}
\setlength{\unitlength}{1cm}
\begin{picture}(8,6.5)
\put(-1.,7.5){\includegraphics{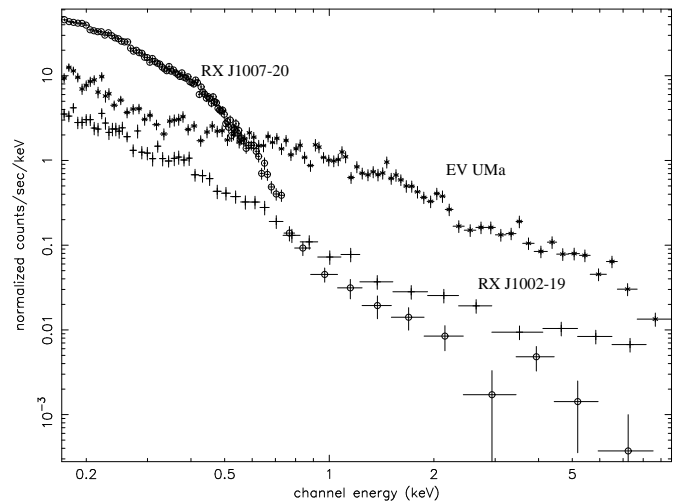}}
\end{picture}
\end{center}
\caption{The X-ray spectra of EV UMa, RX J1002-19 and RX J1007-20.}
\label{spec} 
\end{figure}

We show in Figure \ref{spec} the EPIC pn spectra from all three
polars. As expected from the low hard X-ray count rate of RX J1007--20
(Figure \ref{lightrx1007}) the spectrum of is dominated by a strong
soft X-ray component. Similarly, the strong hard X-ray count rate seen
in the light curve of EV UMa is reflected in a prominent hard
component (it is much stronger than that of RX J1007--20). The
relative strength of the soft and hard components in RX J1002--19 is
midway between the other two.  There maybe some evidence for line
emission near the Fe K$\alpha$ complex in the spectrum of EV UMa but
not in RX J1002--19 or RX J1007--20. The absence of such a feature is
due to the lower flux and hence lower signal-to-noise ratio in these
systems compared to EV UMa.

We define the hard X-ray luminosity as
($L_{hard}=4\pi$Flux$_{hard,bol}d^{2}$) where Flux$_{hard,bol}$ is the
unabsorbed, bolometric flux from the hard component and $d$ is the
distance. Since a fraction of this flux is directed towards the
observer, we switch the reflected component to zero after the final
fit to determine the intrinsic flux from the optically thin post-shock
region (we take into account reflection of hard X-rays from the
surface of the white dwarf in our emission model - {\sl c.f.} next
section).  We define the soft X-ray luminosity as
($L_{soft}=\pi$Flux$_{soft,bol}$sec($\theta)d^{2}$), where we assume
that the soft X-ray emission is optically thick and can be
approximated by a small thin slab of material. The unabsorbed
bolometric flux is Flux$_{soft,bol}$ and $\theta$ is the mean viewing
angle to the the accretion region. Our emission model of the
post-shock region also allows us to estimate the mass of the white
dwarf if we assume the white dwarf mass-radius relationship of
Nauenberg (1972).

The standard accretion shock model of Lamb \& Masters and King \&
Lasota (1979) suggests that the ratio of the flux from the soft X-ray
reprocessed to the post-shock emission should be $\sim$0.5.  These
values will be a lower estimate since emission from cyclotron emission
will contribute to the post-shock luminosity.

\subsection{The model}

We modelled the data using a simple neutral absorber and an emission
model of the kind described by Cropper et al (1999). This emission
model, unlike single temperature thermal bremsstrahlung models, is a
more realistic physical description of the post-shock accretion region
in polars. It is based on the prescription of Aizu (1973) which
predicts the temperature and density profiles over the height of the
accretion shock. However, it has been modified to take into account
cyclotron cooling (which can be significant in polars) and also the
variation in gravitational force over the shock height. 

To reduce the number of free parameters we fix $\epsilon_{s}$, (the
ratio of cyclotron cooling to thermal bremsstrahlung cooling at the
shock front), at a value which implies a magnetic field strength for
the polar is question (the magnetic field strength is a function of
$\epsilon_{s}$, mass transfer rate and mass of the white dwarf - {\sl
c.f.} equation 2 Wu, Chanmugam \& Shaviv 1995).

We also fix the specific accretion rate in the range 1--5 g s$^{-1}$
cm$^{-2}$ (typical of polars in an intermediate to high accretion
state). Changing these parameters does not have a great affect on the
results for data with moderate signal-to-noise ratio. We also include
reflection of hard X-rays from the surface of the white dwarf using
the results of van Teeseling et al (1996). We assume a mean viewing to
the reflecting site of 30$^{\circ}$. We also added a blackbody model
and determined the change in the fit.

\subsection{EV UMa}

\begin{figure}
\begin{center}
\setlength{\unitlength}{1cm}
\begin{picture}(8,5.5)
\put(-0.8,-0.8){\includegraphics{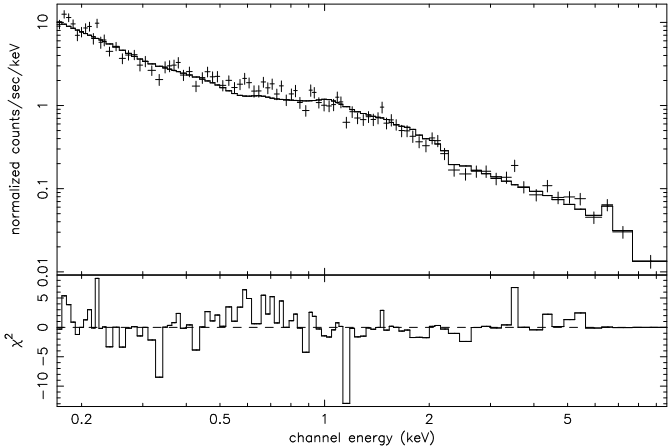}}
\end{picture}
\end{center}
\caption{The EPIC pn spectra of EV UMa together with the best
fit model and the residuals.}
\label{specev} 
\end{figure}

We show the EPIC pn spectrum from $\phi$=0.70--0.95 together with the
residuals in Figure \ref{specev}. Whilst we obtain a reasonably good
fit, there is an excess of residuals around 0.5--0.8keV. This maybe
due to the presence of photoionized line emission which is not
accounted for in our emission model. There is no evidence for a
significant difference between this spectra and the spectra extracted
from $\phi$=0.05--0.35. Further, all EPIC spectra show a low Hydrogen
column density ($<6\times10^{19}$ \pcmsq). The mean viewing angle to
the emission region is uncertain. We initially assume a mean viewing
angle of 50${^\circ}$. We therefore have to apply a correction factor
of sec$\theta$=1.55 to the blackbody luminosity to take into account
that this component is optically thick. EV UMa is one of the most
distant polars known: Osborne et al (1994) put a lower limit to the
distance of $\sim$700pc and is at least 630pc above the galactic
plane. In determining the luminosities we assumed a distance of 700pc.

We find that the mass of the white dwarf is reasonably well
constrained ($M_{wd}\sim$1.0--1.1\Msun). In fitting the spectra from
$\phi$=0.05--0.35 we fix the mass at the values we found from our fits
to the spectra extracted from $\phi$=0.70--0.95. Assuming a viewing
angle of 50${^\circ}$ we find the ratio
$L_{soft}/L_{hard}\sim$0.1--0.2. To increase this ratio so that it
matches that predicted by the standard model (Lamb \& Masters 1979,
King \& Lasota 1979) we would require a mean viewing angle
$\sim80{^\circ}$. The Hakala et al (1994) polarisation data modelling
predicts that the upper accretion region is seen to rotate into and
out from view and hence a high viewing angle is not unreasonable. We
suggest in \S 3.1 that the phase range $\phi$=0.05--0.35 coincides
with the accretion region gradually going behind the limb of the white
dwarfs so the viewing angle to this emission site is expected to be
even higher than for the previous spectrum.
   
\subsection{RX J1002--19}

\begin{figure}
\begin{center}
\setlength{\unitlength}{1cm}
\begin{picture}(8,5.5)
\put(-0.8,-0.8){\includegraphics{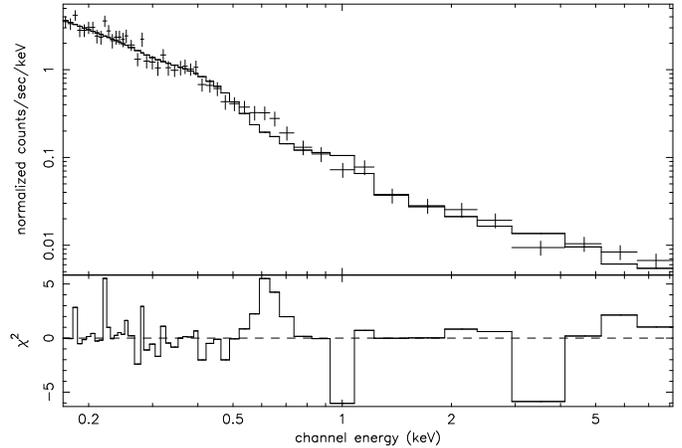}}
\end{picture}
\end{center}
\caption{The EPIC pn spectra of RX J1002--19 together with the best
fit model and the residuals.}
\label{specrx1002} 
\end{figure}

The EPIC pn spectrum together with the residuals are shown in Figure
\ref{specrx1002}. Similarly to EV UMa, we find an excess of residuals
around 0.5--0.7keV. The spectrum of RX J1002--19 is unusual in that it
shows evidence for strong absorption by a neutral absorber with
partial covering, (a simple absorber gave significantly poorer
fits). The viewing angle to the accretion region is uncertain in this
system. However, we do observe it coming over and behind the limb of
the white dwarf so the mean viewing angle to the accretion region must
be relatively high. We assume a mean viewing angle of 50$^{\circ}$
giving a correction factor of 1.55 for the reprocessed component. We
do not know the distance to this system so we assume a distance of
100pc in determining the luminosities.

The resulting mass of the white dwarf is not strongly constrained,
although the EPIC pn spectrum does indicate that it is less massive
than $\sim$1.0\Msun. The MOS spectrum was very poorly constrained so
we fixed it at the best fit found from the pn spectrum (0.5\Msun). In
determining the flux of the shocked component we keep the mass
fixed. The resulting ratio $L_{soft}/L_{hard}$ ($\sim$1--2) is
slightly greater than expected from the standard shock model. However,
if there is a significant cyclotron emission present then this would
add to the flux from the shocked component. It would therefore result
in a lower ratio which would be close to that expected from the
standard shock model.

\subsection{RX J1007--20}

The spectrum of RX J1007--20 is dominated by a strong soft X-ray
component.  Although the resulting estimate for the mass of the white
dwarf is not well constrained, a mass of $\sim$1\Msun is consistent
with both the MOS and pn data. When we determine the flux from the
shock component we fix the mass at the best fit value. The viewing
angle to the accretion region is uncertain. Therefore, we assume a
rather conservative estimate for the viewing angle of 30$^{\circ}$
giving a correction factor for the reprocessed component of 1.15. In
determining the luminosity we assume a distance of 700pc (Reinsch et
al 1999).

The resulting ratio, $L_{soft}/L_{hard}$ is $\sim$8 and $\sim$11
(determined from the pn and MOS detectors respectively) is high:
clearly RX J1007-20 has a large `soft X-ray excess'.  The light curve
is highly variable: this is consistent with the view that the soft
X-ray excess is due to dense blobs of material which impact the
photosphere of the white dwarf directly causing their energy to be
re-processed as soft X-rays (Kuijpers \& Pringle 1982).

\begin{figure}
\begin{center}
\setlength{\unitlength}{1cm}
\begin{picture}(8,5.5)
\put(-0.8,-0.3){\includegraphics{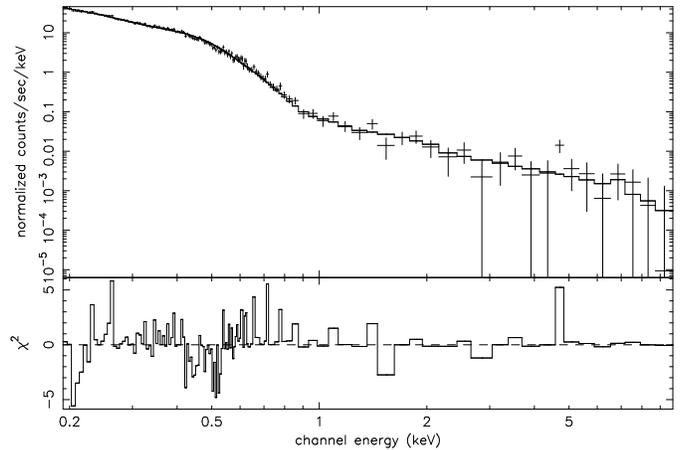}}
\end{picture}
\end{center}
\caption{The EPIC pn spectra of RX J1007--20 together with the best
fit model and the residuals.}
\label{specrx1007} 
\end{figure}

\begin{table*}
\begin{center}
\begin{tabular}{llrrrrrrrrr}
\hline 
Source & Detector & $N_{H} (10^{20}$ & $kT_{bb}$ & Flux &
$L_{bb,bol}$ & Flux & $L_{hard,bol}$ & 
$L_{bb,bol}/$ &
$M_{wd}$ & \rchi\\ 
& & \pcmsq) & (eV) & (bb) &  &
(Shock) & & $L_{hard,bol}$ & (\Msun) & (dof) \\ 
\hline 
EV UMa (1) & PN & 0.0$^{+0.6}$ & 48$^{+4}_{-6}$ &
4.3$^{+1.7}_{-0.4}$ & 98$^{+39}_{-10}$ & 
13$^{+4}_{-7}$ & 735$^{+245}_{-196}$ &
0.13$^{+0.12}_{-0.04}$& 1.01$^{+0.09}_{-0.08}$& 1.82 (89) \\
EV UMa (1) & MOS & 0.0$^{+0.6}$ & 49$^{+7}_{-5}$ &
3.9$^{+1.4}_{-0.8}$ & 93$^{+29}_{-5}$ &
14$^{+6}_{-1}$ & 784$^{+343}_{-49}$ &
0.12$^{+0.05}_{-0.06}$& 1.09$^{+0.07}_{-0.11}$& 1.12 (102) \\
EV UMa (2) &  PN & 0.0$^{+0.3}$ & 40$^{+4}_{-4}$ &
6.4$^{+2.4}_{-1.1}$ & 142$^{+54}_{-25}$ & 
10$^{+0.4}_{-0.4}$ & 539$^{+49}_{-49}$ &
0.26$^{+0.14}_{-0.06}$& 1.01 (fix) & 1.63 (73) \\
EV UMa (2) & MOS  & 0.0$^{+0.5}$ & 56$^{+4}_{-8}$ &
3.0$^{+1.0}_{-0.4}$ & 64$^{+25}_{-10}$ & 
10$^{+2}_{-2}$ & 588$^{+147}_{-98}$ &
0.11$^{+0.08}_{-0.06}$& 1.09 (fix) & 1.30 (139) \\
RX J1002--19 & PN & 480$^{+270}_{-130}$ & 59$^{+3}_{-5}$ &
9.4$^{+1.6}_{-1.3}$ & 4.3$^{+0.8}_{-0.6}$ & 
2.3$^{+0.7}_{-0.5}$ & 2.7$^{+0.8}_{-0.5}$ &
1.6$^{+0.8}_{-0.5}$& 0.5$^{+0.4}_{-0.1}$ & 1.41 (45) \\
& & cf=0.84$^{+0.02}_{-0.03}$ & & &  &  &  & & & \\
RX J1002--19 & MOS & 330$^{+270}_{-120}$ & 58$^{+7}_{-8}$ &
5.9$^{+8.3}_{-0.9}$ & 2.8$^{+3.8}_{-0.9}$ & 
1.5$^{+0.6}_{-0.4}$ & 1.8$^{+0.7}_{-0.5}$ &
1.6$^{+3.6}_{-0.8}$& 0.5 (fix) & 0.53 (31) \\
& & cf=0.76$^{+0.06}_{-0.07}$ & & &  &  &  & & & \\
RX J1007--20 & PN & 1.10$^{+0.25}_{-0.25}$ & 61.5$^{+2.5}_{-2.0}$ &
22$^{+1}_{-2}$ & 372$^{+5}_{-34}$ & 
0.8$^{+0.2}_{-0.1}$ & 47$^{+12}_{-6}$ &
8.1$^{+1.4}_{-2.1}$& 1.04$^{+0.20}_{-0.32}$ & 1.17 (125) \\
RX J1007--20 & MOS & 1.3$^{+0.4}_{-0.6}$ & 55$^{+2}_{-2}$ &
32$^{+5}_{-7}$ & 540$^{+50}_{-100}$ & 
0.8$^{+0.1}_{-0.1}$ & 45$^{+3}_{-5}$ &
11.8$^{+2.8}_{-2.8}$& 1.05$_{-0.12}$ & 1.64 (90) \\
\hline
\end{tabular}
\end{center}
\caption{The fits to the X-ray data. EV UMa (1) and EV UMa (2) refer
to the bright phase before and after the absorption dip -- see text
for details. For RX J1002-19 we have used a partial covering absorber:
`cf' refers to the covering fraction. The blackbody bolometric
luminosity, $L_{bb,bol}$ and the X-ray luminosity from post-shock
flow, $L_{hard}$, are defined in the text. The units of flux are
$10^{-12}$ \ergscm and luminosity are $10^{30}$ \ergss. We assume a
distance to RX J1007--20 and EV UMa of 700pc and 100pc to RX
J1002--19.}
\label{fits}
\end{table*}

\section{Discussion}

We have presented {\xmm} data of three polars, which display rather
different characteristics. We find that in each system we cannot
obtain good fits to their X-ray spectra using a single component model
-- they all require a shock model plus a soft blackbody
component. However, all three show very different soft/hard X-ray
ratios. Further, their light curves show a range of features. We now
go on to discuss our findings.

\subsection{EV UMa - one or two accretion poles?}

The X-ray light curves and the hardness ratio of EV UMa (Figure
\ref{lightev}) suggested a different accretion region may dominate
before and after the absorption dip. By examining the X-ray spectra
from these phase ranges we find that there is no evidence that their
spectra differ. Together with the phasing of the X-ray light curves
and the optical light curves (\S \ref{evphase}) we suggest that there
is only one dominant accretion region and the presence of negative
circular polarisation for a short time in the optical data of Hakala et
al (1994) is due to observing the cyclotron emitting region from
beneath.

\subsection{Soft/Hard ratio}

EV UMa, RX J1002-19 and RX J1007-20 show very different
$L_{soft}/L_{hard}$ ratios, with EV UMa showing the lowest and RX
J1007--20 the highest. Indeed, RX J1007--20 has a ratio which is a
factor of $\sim$20 greater than that predicted from the `standard'
shock model of Lamb \& Masters (1979) and King \& Lasota
(1979). However, the fact that it has a high magnetic field strength
(92 MG, Reinsch et al 1999), a large `soft X-ray excess' is consistent
with the work of Ramsay et al (1994) and Beuermann \& Burwitz (1995)
who showed that this ratio was correlated with magnetic field
strength; with high field systems showing high soft/hard ratios. To
account for this, the most likely solution is that dense blobs of
material impact with the white dwarf directly without forming a shock
(Kuijpers \& Pringle 1982) or the shock is buried sufficiently deep
for the bremsstrahlung emission to be thermalized in the local
photosphere of the white dwarf (Frank, King \& Lasota 1988). The fact
that we observe strong flaring activity in the light curve of RX
J1007--20 is consistent with this view.

Based on the soft/hard ratio we predict that RX J1002-19 will have a
magnetic field strength which is slightly greater than that of EV
UMa. If we observe the accretion region(s) of EV UMa at a high viewing
angle then its soft/hard ratio is consistent with the standard shock
model. 

Ramsay et al (2001) showed that using a single temperature thermal
bremsstrahlung model for the hard X-ray component rather than a
stratified shock as used here, the resulting ratio $L_{soft}/L{hard}$
is higher: in the case of DP Leo by a factor of 2. This should be
taken into account when comparing with studies using single
temperature models.

\subsection{Masses}

We can infer the mass of the white dwarf from our model fitting
assuming a mass-radius relationship for the white dwarf. Using the
Nauenberg (1972) relationship we find that both EV UMa and RX
J1007--20 have masses of $\sim1$\Msun. In the case of RX J1002--19 the
mass is not very well constrained, although it is likely to be less
than 1\Msun with a best fit of 0.5\Msun.

Using the same model for the shock region as here, Ramsay (2000)
fitted {\sl RXTE} spectra from 21 mCVs and found that their masses
were biased towards higher masses than that of isolated white
dwarfs. There was no significant difference between the mass of the
white dwarf in mCVs and non-magnetic CVs. This is consistent with the
masses reported here and also the other mCVs which have been observed
using {\xmm}: CE Gru ($\sim$1.0\Msun, Ramsay \& Cropper 2002b), BY Cam
(0.9--1.1\Msun, Ramsay \& Cropper 2002a), DP Leo ($>$1.3\Msun, Ramsay
et al 2001) and WW Hor ($\sim$1.0--1.1\Msun, Ramsay et al 2001). It
remains to be seen whether there is a systematic bias in the masses
determined with this method using {\xmm} data.

\subsection{Mass transfer rate}
\label{transfer}

To determine the mass transfer rate we use the standard relation,
$L_{acc}=GM_{wd}\dot{M}/R_{wd}$. The accretion luminosity, $L_{acc}$,
is the sum of $L_{hard}$, the unreprocessed fraction of the cyclotron
luminosity, $L_{cyc}$, and the luminosity of any dense blobs of
material which do not form a shock and emit in soft X-rays. We use the
best fit masses determined from our model fits to determine the mass
accretion rate for EV UMa and RX J1007--20 (we omit RX J1002--19 since
there is no estimate for its distance).

In the case of EV UMa, the results of our spectral analysis indicate
there is no evidence for blobby accretion. Using the model results of
Woelk \& Beuermann (1996) and assuming a magnetic field strength of
(30-40MG) and a mass $M_{wd}$=1.0\Msun, we find
$L_{cyc}/L_{hard}\sim$1--10 for a range of $\dot{m}$=1--10 g cm$^{-2}$
s$^{-1}$. This implies $L_{cyc}=7.5-70\times10^{32}$ \ergss. The mass
transfer rate is therefore $\sim6-30\times10^{15}$ g s$^{-1}$ and
hence the fraction of the white dwarf that is accreting is
$\sim1.6-8.3\times10^{-3}$.

In the case of RX J1007--20 the spectral results indicate that a
significant proportion of the soft X-ray luminosity is in the form of
blobby accretion. We therefore include $L_{soft}$ when determining
the total luminosity. We again use the model results of Woelk \&
Beuermann (1996) to estimate $L_{cyc}$ and assume $B=$92MG and
$M_{wd}$=1.0\Msun. We find that $L_{cyc}\sim10-100\times
(L_{hard}+L_{soft})$ and hence $L_{cyc}\sim4-50\times10^{33}$
\ergss. This implies $\dot{M}\sim2-20\times10^{16}$ g s$^{-1}$ and a
fractional area of $\sim5-60\times10^{-3}$.

It is difficult to compare these mass transfer rates with previous
results because of the different way authors account for the cyclotron
flux and whether they include the soft X-ray luminosity. However, the
fractional area of the white dwarf which is accreting is consistent
with previous studies.

\subsection{Absorption dips}

\begin{figure}
\begin{center}
\setlength{\unitlength}{1cm}
\begin{picture}(8,10)
\put(-0.5,-1.4){\includegraphics{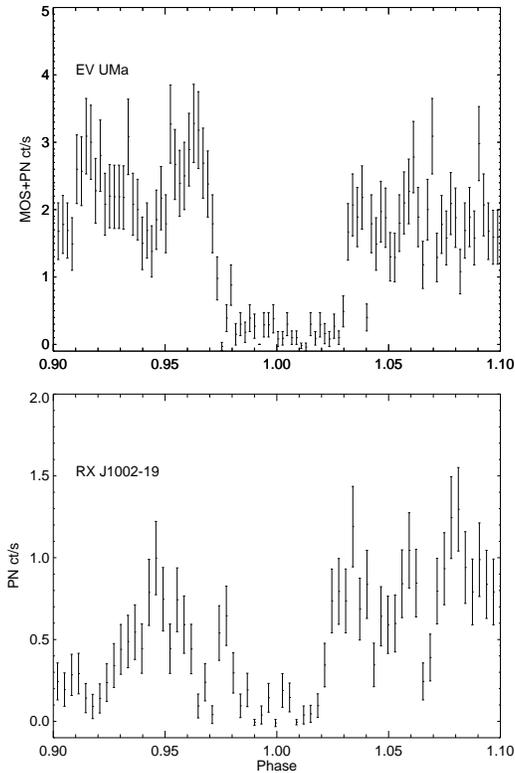}}
\end{picture}
\end{center}
\caption{The EPIC pn X-ray light curves (0.15--1.0keV) of EV UMa and
RX J1002-19 showing the accretion stream dip in detail. The
integration time is 10 and 20 sec for EV UMa and RX J1002-19
respectively. We have phased the light curves so that phase 0.0
corresponds to the centre of the dip.}
\label{dip} 
\end{figure}

Absorption dips of the kind seen in EV UMa and RX J1002--19 are caused
by the accretion stream obscuring the emission sites on the white
dwarf as it passes through our line of sight. For an accretion region
in the upper hemisphere of the white dwarf, such a dip is inevitable
if the latitude of the region, $m$, is less than the binary
inclination, $i$. In the case of EV UMa, Hakala et al (1994) finds the
optical polarisation data is best modelled with a high inclination
($\sim75^{\circ}$) so it is likely that this condition holds. The
inclination for J1002--19 is currently unknown. The dip ingress and
egress in EV UMa is rapid: the ingress takes 20--30 sec and less than
20 sec for the egress (Figure \ref{dip}). Such a rapid ingress and
egress imply that the stream is highly collimated. For RX J1002--19 the
dip structure is more complex (Figure \ref{dip}) with a shorter dip
preceding the main dip which also has a sharp profile.

The dip duration (defined as the full width half maximum of the dip
profile) is 0.057 and 0.042 cycles for EV UMa and RX J1002--19
respectively (where we do not include the shorter dip). Using equation
(14) of Watson et al (1989) we find that for the observed duration of
the dips, $r_{s}=0.18d$ and 0.13$d$ for EV UMa and RX J1002--19
respectively where $r_{s}$ is the radius of the accretion stream and
$d$ is the distance between the white dwarf and the source which is
causing the absorption dip.  (We assume the stream is in the orbital
plane). For a stream radius of 10$^9$ cm these radii imply
$d=5.5\times10^9$ and 7.7$\times10^9$ cm.

To make a very approximate estimate of the radius at which the
accretion stream becomes coupled by the magnetic field, $R_{\mu}$, we
use equation (1b) of Mukai (1988). If we use the best fit masses, the
mass accretion rates determined in \S \ref{transfer}, the radius of
the accretion stream, $r=1\times10^{9}$ cm, and $B$=30MG we derive
$R_{\mu}=1.1\times10^{10}$ and 3.5$\times10^{10}$ cm for EV UMa and RX
J1002--19. Although there is a great deal of uncertainty in some of
the values of these parameters and the applicability of equation (1b)
of Mukai (1988), it does suggest that the stream which is obscuring
the accretion region during the dip is located in the magnetically
controlled portion of the accretion flow.

We can make an estimate of the total column density of the stream by
taking our best model fits then increasing the absorption until the
model count rate matches the observed count rate at dip maximum for
the specified energy range. We find a column density of
7--20$\times10^{22}$ \pcmsq and $\sim1\times10^{21}$ \pcmsq for EV UMa
and RX J1002--19 respectively. We can make an estimate of the number
density of the stream using equation (6) of Watson et al
(1995). Assuming the accretion flow is in the orbital plane and using
the above values of the total column density we find $n_{13} r_{9}$=5
and 0.05 for EV UMa and RX J1002--19 respectively, where $n_{13}$ is
the constant number density in units of 10$^{13}$ cm$^{-3}$ and
$r_{9}$ is the radius of the stream in units of 10$^{9}$ cm. For
comparison Watson et al (1995) found $n_{13}r_{9}$=5 for RX
J1940-10. Taking these values at face value, they imply that either
the accretion stream in RX J1002--19 has a rather small radius or a
low number density.

\section{acknowledgments}

We acknowledge the use of the {\ros} all-sky survey data archive held at
MPE, Garching, Germany.

\end{document}